\documentclass[A4,titlepage,11pt]{article}

\pdfoutput=1

\usepackage{amssymb,amsmath,amsfonts}
\usepackage{ccaption}
\usepackage{graphicx}
\usepackage{subfigure}
\usepackage{caption}

\usepackage[
      colorlinks=true,
      linkcolor=blue,
      urlcolor=blue,
      filecolor=black,
      citecolor=red,
      pdfstartview=FitV,
      pdftitle={NSBH Merger},
        pdfauthor={Roberto Emparan, Daniel Mar{\'\i}n},
        pdfsubject={},
        pdfkeywords={},
        pdfpagemode=None,
        bookmarksopen=true
      ]{hyperref}

\def\clock{{\count0=\time
           \divide\count0 60
           \ifnum\count0<10 0\fi\the\count0
           \multiply\count0 -60 \advance\count0 \time
           :\ifnum\count0<10 0\fi \the\count0
         }}
\newcommand{\timestamp}{{\small\vbox{\hbox{\tt\jobname.tex}
\hbox{\the\day/\the\month/\the\year, \clock}}}}

\newcommand{\eg}{{\it e.g.,\ }}

\newcommand{\lp}{\left(}
\newcommand{\rp}{\right)}

\newcommand{\beq}{\begin{equation}}
\newcommand{\eeq}{\end{equation}}
\newcommand{\bea}{\begin{eqnarray}}
\newcommand{\eea}{\end{eqnarray}}
\newcommand{\beqa}{\begin{eqnarray}}
\newcommand{\eeqa}{\end{eqnarray}}

\numberwithin{equation}{section}

\begin{document}

\begin{titlepage}
\leftline{}
\vskip 2cm
\centerline{\LARGE \bf Precursory collapse in} 
\bigskip
\centerline{\LARGE \bf Neutron Star\,-\,Black Hole mergers}
\vskip 1.2cm
\centerline{\bf Roberto Emparan$^{a,b}$, Daniel Mar{\'\i}n$^{b}$}
\vskip 0.5cm
\centerline{\sl $^{a}$Instituci\'o Catalana de Recerca i Estudis
Avan\c cats (ICREA)}
\centerline{\sl Passeig Llu\'{\i}s Companys 23, E-08010 Barcelona, Spain}
\smallskip
\centerline{\sl $^{b}$Departament de F{\'\i}sica Qu\`antica i Astrof\'{\i}sica, Institut de
Ci\`encies del Cosmos,}
\centerline{\sl  Universitat de
Barcelona, Mart\'{\i} i Franqu\`es 1, E-08028 Barcelona, Spain}
\smallskip

\vskip 1.2cm
\centerline{\bf Abstract} \vskip 0.2cm 
\noindent 
We investigate the properties of the event horizon in the merger between a large black hole and a smaller neutron star. We find that, if the star is compact enough, then, in its rest frame a horizon begins to grow inside the star before it merges with the black hole, in a manner analogous to the growth of the event horizon in stellar collapse. We may say that, ahead of its fall into the larger black hole, the neutron star begins to become a black hole itself. We discuss how the phenomenon, even if not directly observable, can be invariantly characterized.  We demonstrate it quantitatively by explicitly constructing the merger event horizon in the extreme-mass-ratio limit. We show that the effect is present for realistic neutron star models and admissible values of the compactness.

\end{titlepage}
\pagestyle{empty}
\small
\normalsize
\newpage
\pagestyle{plain}
\setcounter{page}{1}

\section{Introduction and Summary}

By now, gravitational wave observatories have detected a variety of mergers between the most compact objects in nature---black holes and neutron stars---in highly dynamical events where the geometry of spacetime is pushed to its limits \cite{ligo}.
In theoretical studies, these are simulated by evolving the Einstein equations, possibly with a suitable matter model for the star, with a focus on extracting waveform templates for the radiation that is emitted. Even though this attention on the signal that is directly measurable in the detectors is understandable, there are other less well studied aspects of these mergers that can teach us about spacetime distortion under extreme conditions.

In this article we explore the evolving features of the event horizon when a large black hole and a smaller neutron star (or, for that matter, a very compact material object) collide and merge. We identify a phenomenon that we refer to as ``precursory collapse'': inside a sufficiently compact star, a horizon begins to grow before it merges with the black hole in the star's rest frame.  We may say that the neutron star, in anticipation of being engulfed by a large black hole, starts becoming a black hole itself. Put differently, the star merges with the black hole from the inside out.

We hasten to emphasize that this horizon growth in the star is not the result of matter being increasingly compressed---indeed, in its rest frame the star is very approximately static. Instead, we will see that it is a consequence of a peculiarity of the event horizon when the infalling star is compact enough. Interestingly, the compactness required is within currently acceptable bounds for physical neutron stars. 

Recall that the event horizon of a black hole makes precise the idea of a region from which nothing can ever escape outside. It is identified as the boundary of what can be seen when looking back in time from an asymptotic region ($\mathcal{I}^+$) at arbitrarily late times. Sometimes this retrospective finality is regarded as unphysical, but this is not the case. The event horizon is an invariant construct, and once a system has reached a state that is close enough to stationarity, over a region that can be approximated as asymptotically flat, then it makes perfect physical sense to reconstruct the event horizon in the spacetime up to that moment. It is true, though, that its features may be counterintuitive, appearing to move in anticipation to the future infall of objects. The effect that we describe may be placed in this class, but it is more striking and suggestive than previously known instances. 

An important consideration for making sense of the appearance of the horizon within the star is the choice of an appropriate reference frame. We will describe the phenomenon in the star's rest frame, which, as we will see, is well defined when the star is much smaller than the black hole. In this reliance on a particular class of time-slicings, the precursory collapse of the star is akin to the observation in \cite{Hughes:1994ea,Shapiro:1995rr} of transient toroidal sections of the event horizon in the collapse to a Kerr black hole and in rotating black hole mergers \cite{Cohen:2011cf,Bohn:2016soe,Emparan:2017vyp}. As in that case, the slice dependence can be interpreted as saying that the phenomenon happens faster than the speed of light: the torus hole closes up superluminally fast, and here also the horizon within the star approaches and merges with the larger black hole horizon along a spacelike trajectory. Obviously, this makes the effect directly unobservable\footnote{Like, in fact, the event horizon is itself, notwithstanding the existence of the so-called Event Horizon Telescope!}. What we find significant is that it is present when there exists a well-defined preferred time direction, namely the star's proper time---so the effect can be invariantly defined---and moreover it happens for physically acceptable compactness. This qualifies it as a neat new property of the event horizon in highly dynamical situations.

The essence of the phenomenon can be easily described in general pictorial terms, and so we will begin in the next section with a qualitative explanation of it. Afterwards, in sec.~\ref{sec:nsbhemr}, we demonstrate the effect with explicit computations of the event horizon in the merger between a large black hole and a much smaller neutron star. For the latter, we study the spherically symmetric Tolman VII solution (T-VII) \cite{Tolman:1939jz} and the Schwarzschild interior solution \cite{Schwarzschild:1916ae}. T-VII is especially relevant since it is a reasonably realistic model \cite{Delgaty:1998uy,Lattimer:2000nx,Raghoonundun:2015wga} that is also explicit enough for detailed calculations. We will study the fusion in the extreme-mass-ratio (EMR) limit where the ratio of neutron star mass to black hole mass is vanishingly small. As shown in \cite{Emparan:2016ylg}, in this case the event horizon of the merger can be constructed very accurately and very simply, and the notion of the star's rest frame is exact and invariant. Moreover, the EMR can be a good approximation to some detections by LIGO-Virgo---currently, several marginal candidate events for NS-BH mergers in runs O1-O3\footnote{For instance, event 151019 in O1 \cite{Nitz:2018imz} and possibly S190814bv and others in O3 \cite{Ackley:2020qkz,Kawaguchi:2020osi}.} have mass ratios $\lesssim 1/10$---and also by upcoming observatories. Many LISA events are indeed expected to fall in this category \cite{Babak:2017tow}.

Our results are easily summarized. The compactness of a neutron star of mass $M$ and radius $R$ is usually characterized with the dimensionless ratio
\beq\label{defbeta}
\beta =\frac{GM}{c^2 R}\,.
\eeq
An absolute upper limit is set by the  Schwarzschild black hole, $\beta_{\textrm{Schw}} = 1/2$, and realistic values for neutron stars lie in the range $\beta \approx 0.18 - 0.25$. For a T-VII star we find that the precursory collapse in the star's rest frame happens whenever
\beq\label{betaT-VII}
\beta > \beta_{\textrm{T-VII}} \equiv 0.22\,.
\eeq
We shall also argue that it occurs in a model-independent manner when
\beq\label{betas}
\beta> 0.2840\,.
\eeq
Even though this may be too large for neutron stars, it applies to exotic compact objects (ECOs) that comfortably comply with the Buchdahl bound (for perfect-fluid stars with radially non-increasing density and isotropic, positive pressure that is finite at the center \cite{Buchdahl:1959zz})
\beq\label{Buchbound}
\beta< \beta_{\textrm{B}}\equiv \frac49=0.44\,.
\eeq

\begin{figure}[htbp]
\centering
\subfigure[$t=-5$]{\includegraphics[width=.32\textwidth]{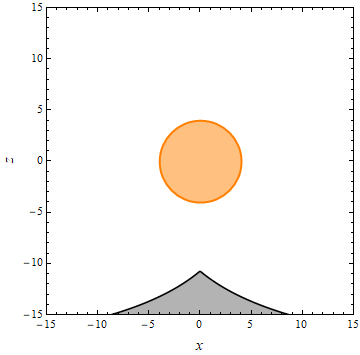}}
\subfigure[$t=-1.3$]{\includegraphics[width=.32\textwidth]{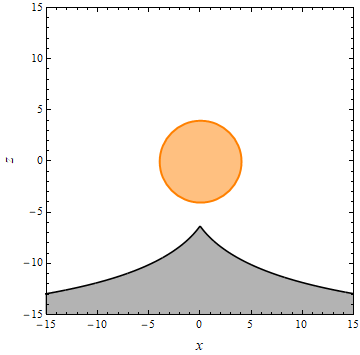}}
\subfigure[$t=-1$]{\includegraphics[width=.32\textwidth]{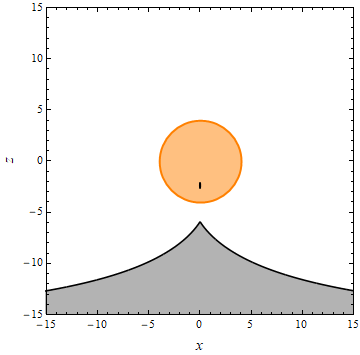}}
\subfigure[$t=-0.3$]{\includegraphics[width=.32\textwidth]{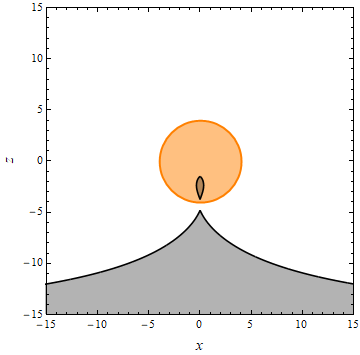}}
\subfigure[$t=0$]{\includegraphics[width=.32\textwidth]{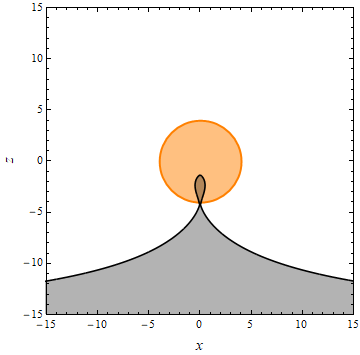}}
\subfigure[$t=0.5$]{\includegraphics[width=.32\textwidth]{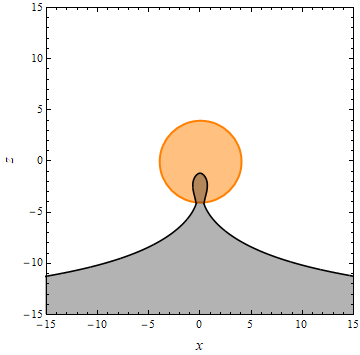}}
\subfigure[$t=1.5$]{\includegraphics[width=.32\textwidth]{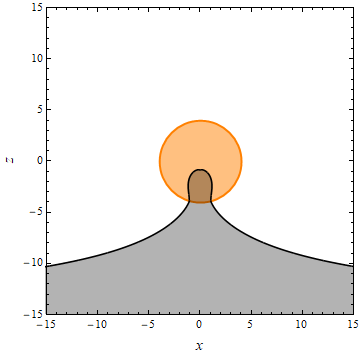}}
\subfigure[$t=7$]{\includegraphics[width=.32\textwidth]{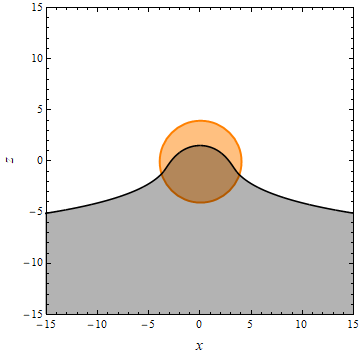}}
\subfigure[$t=11$]{\includegraphics[width=.32\textwidth]{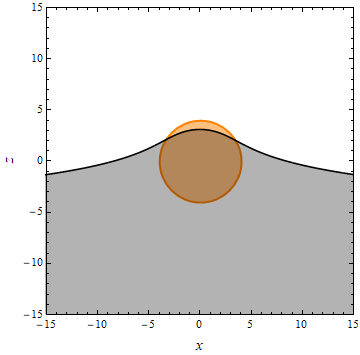}}\caption{\small Sequence of constant-time slices of the event horizon of NS-BH merger with star compactness $\beta=0.25$. Frames are centered on the neutron star (orange disk) of radius $R=4$, in units where $M=1$. The large black hole lies at the bottom. The event horizon is the black line, and the gray-shaded regions are its interior. $t$ is the (Killing) proper time of the star. The precursory collapse, in which the horizon grows inside the star, begins at $t= -1.06905$. The two horizons fuse (e) at $t=0$. The complete constant-time slices are obtained by rotating around the axis $x=0$.} \label{fig:cuts}
\end{figure}

The precursory collapse of the star is most clearly illustrated in a sequence of constant time slices of the event horizon. Figure~\ref{fig:cuts} shows it for the EMR merger of a T-VII star with $\beta=0.25$.

\section{Qualitative picture}\label{sec:qualia}

\begin{figure}
\centering
\subfigure[BH-BH]{\includegraphics[width=.47\textwidth]{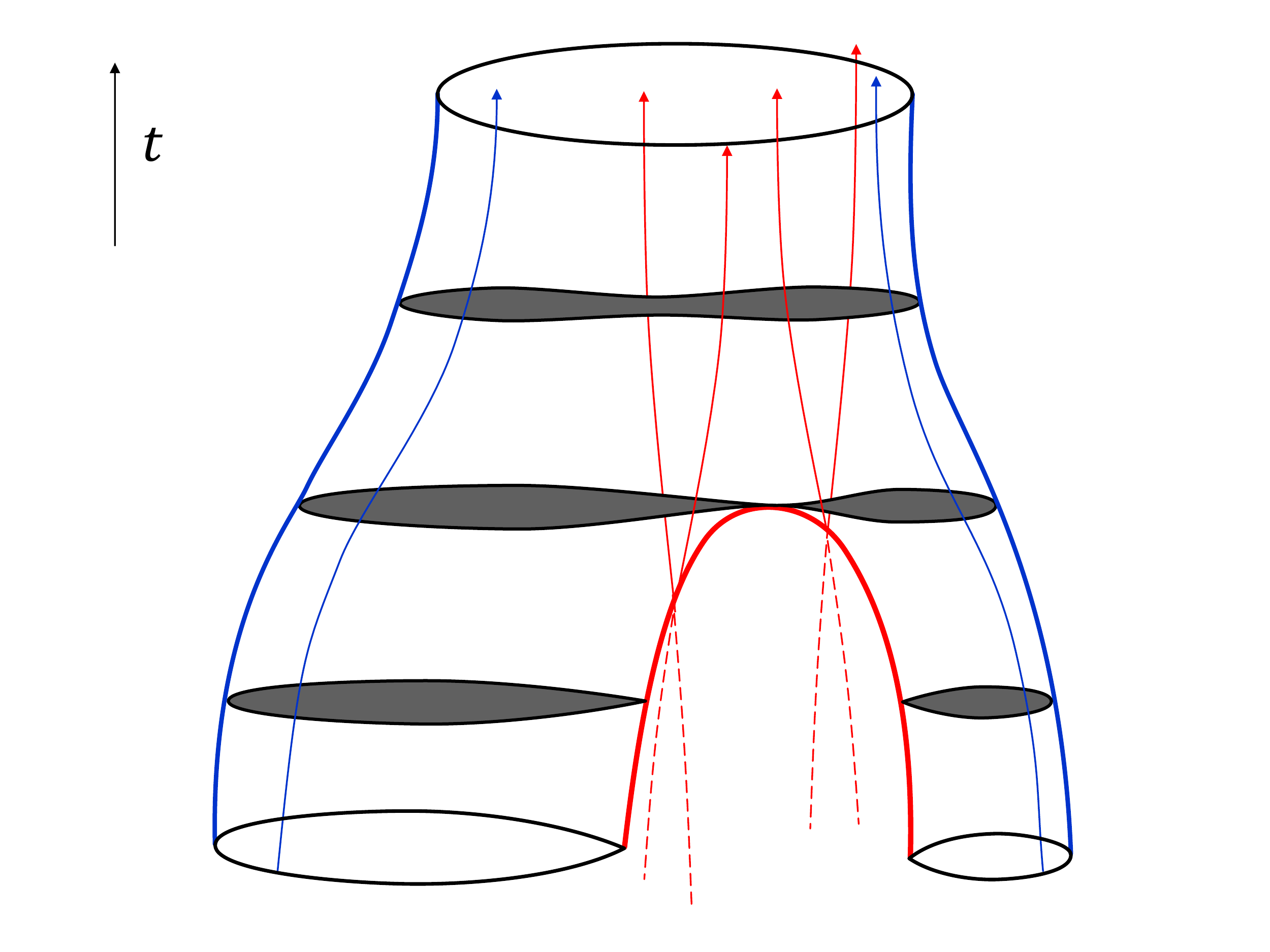}	  \label{fig:BHBH}}
\subfigure[NS-BH, low $\beta$]{\includegraphics[width=.47\textwidth]{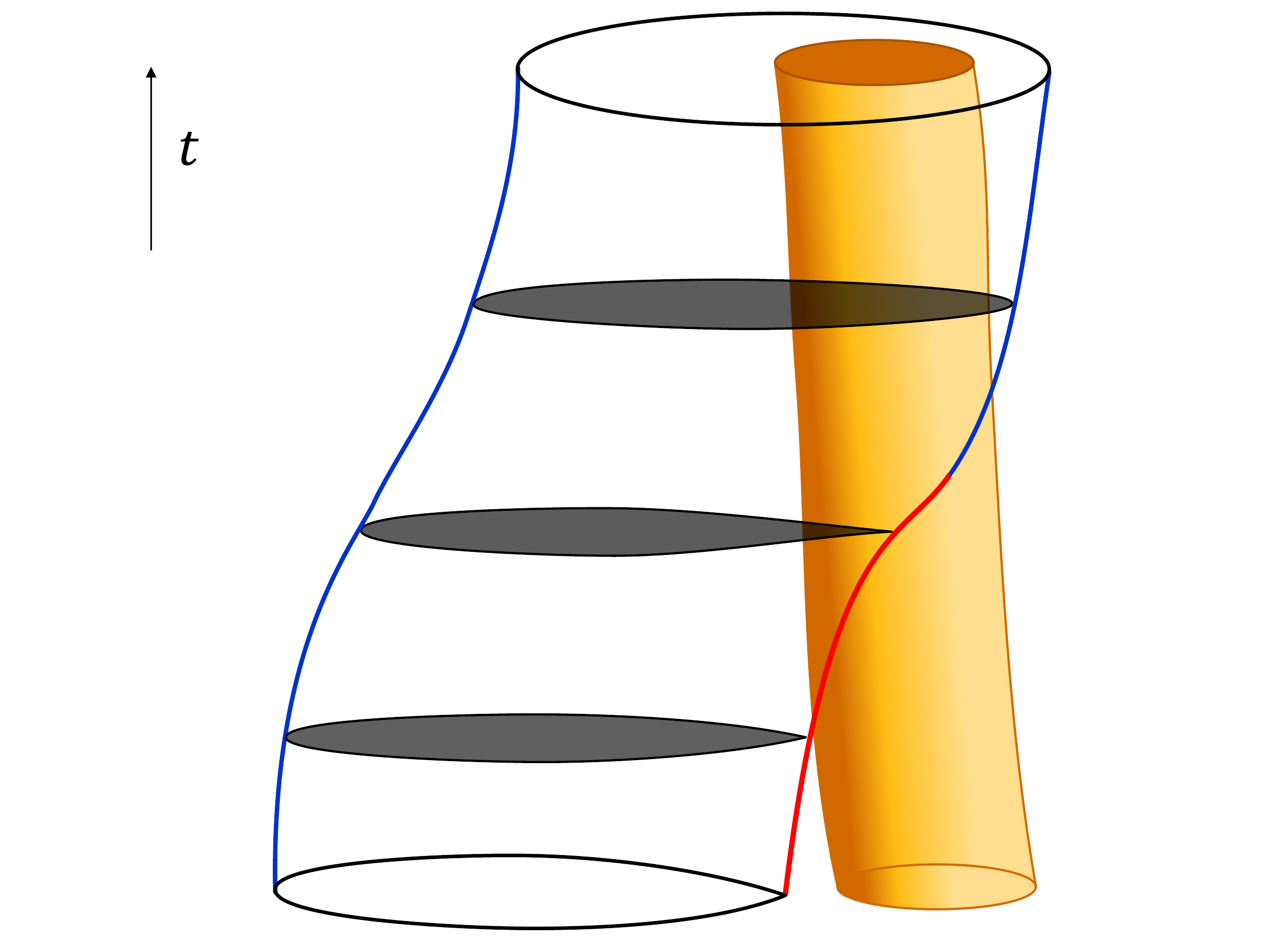} \label{fig:NSBHlo}}
\subfigure[NS-BH, high $\beta$]{\includegraphics[width=.6\textwidth]{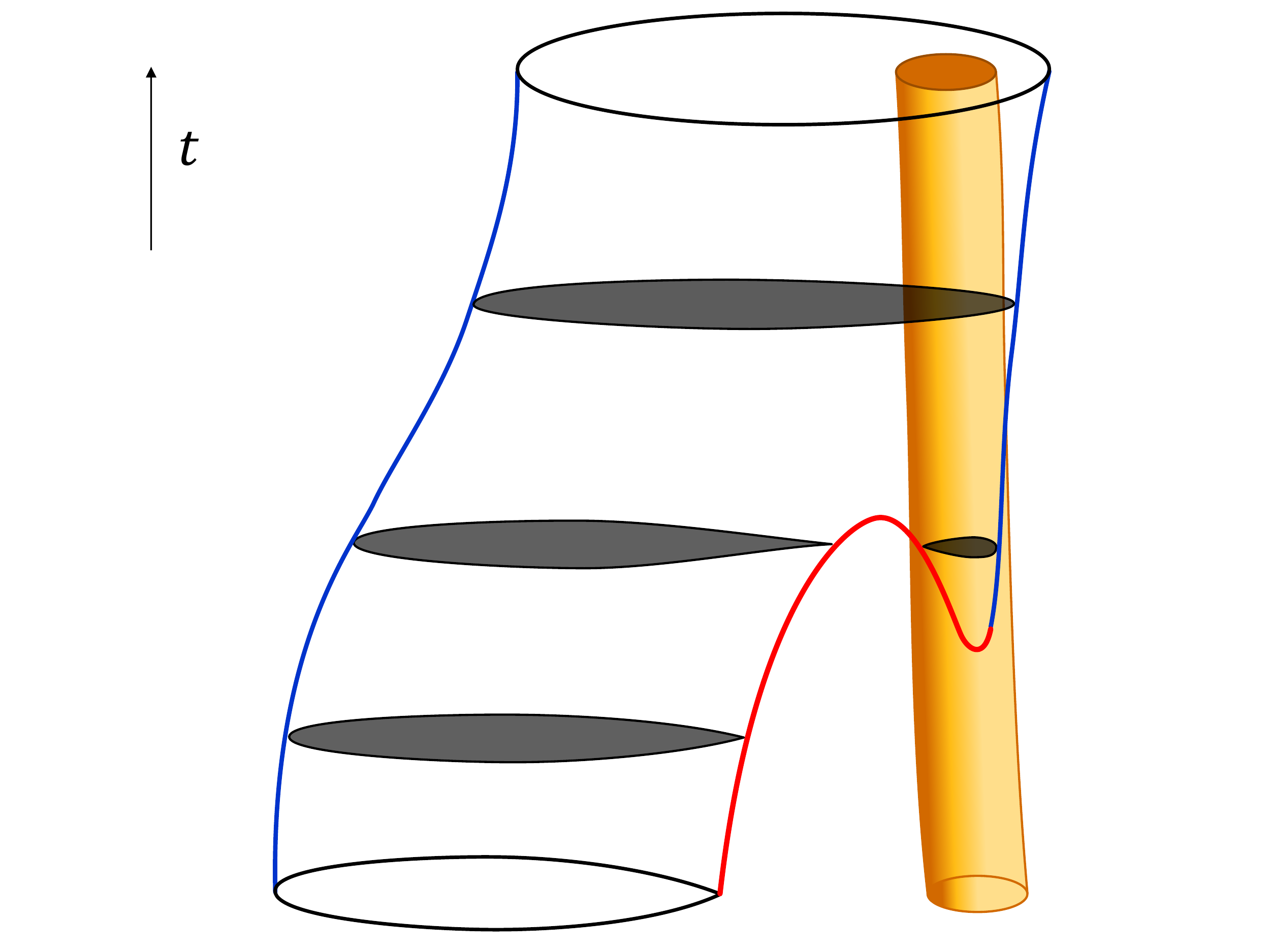}  \label{fig:NSBHhi}}

\caption{\small Event horizon in Binary Black Hole and Neutron Star-Black Hole mergers. Spatial sections of the horizon are gray-shaded. In (a) we draw some light rays that generate the horizon; red rays enter the horizon at the spacelike caustic (thick red). In (b) and (c) the orange world-tube is the neutron star, for low and high compactness $\beta=M/R$. When the star is compact enough, (c), the spatial cross sections (in the rest frame of the star, or close to it) develop a component of the horizon inside the star---the ``precursory collapse''---which then merges with the black hole.}
\label{fig:mergers}
\end{figure}

Let us begin by recalling the main features of the event horizon of a binary black hole merger, which generically takes the form of the `pants' diagram in figure~\ref{fig:BHBH}. The surface is generated by null rays and is smooth except at the places where new null generators enter towards the future to form part of the horizon. This is the `crease set', which is a spacelike point set in an otherwise null surface where null rays focus (caustic points) or cross each other (crossover points) as they enter the future horizon \cite{Lehner:1998gu,Husa:1999nm,Siino:2004xe,Siino:1997ix,Siino:2005nq}. In a merger, the crease lies at the crotch of the pants. In axisymmetric collisions it is a line of caustics along the collision axis; this case has been thoroughly characterized  in \cite{Hamerly:2010cr,Emparan:2016ylg,Emparan:2017vyp,Hussain:2017ihw}.

The event horizon is a 3-surface, but we are interested in following the evolution of the merger along sequences of two-dimensional spacelike sections. In binary black hole mergers, these sequences show two horizon components approaching and fusing into a single one.

Replace now the smaller black hole in the binary with a neutron star. The thinner leg of the pants disappears, but there remains a portion of the crease where new null rays are added to the horizon as the star approaches the black hole. When taking spacelike sections, in general there is no preferred temporal slicing, but when the star is much smaller than the black hole the tidal effects on it are very small and it retains its shape throughout the merger. The equivalence principle tells us that, in its free fall into its gargantuan companion, the star is essentially unaffected. Therefore in this limit we can very approximately define the rest frame of the star, and choose its proper time $t$ as our time coordinate.

If the star is not very compact, with $\beta$ quite smaller than that of the black hole, the region amputated from the pants of figure~\ref{fig:BHBH} is quite larger than the thin leg. This results in a diagram like figure~\ref{fig:NSBHlo}. The sections of the event horizon have a single horizon component. However, if the star is more compact, less is amputated from the horizon around the thinner leg, and the cap of the crease, with a $\cap$ shape (in axisymmetric collisions it is actually a saddle), can remain, see figure~\ref{fig:NSBHhi}. Taking slices at constant $t$, during some time the horizon has two components, one of them inside the star, which grows from zero size in a manner analogous to the horizon inside a collapsing star, until it merges with the larger black hole horizon.

It should be clear that, since the crease set is spacelike, one may always choose slices where the horizon sections have a single component as in figure~\ref{fig:NSBHlo}. Conversely, an event horizon like the one in figure~\ref{fig:NSBHlo} can always be sliced in such a way that two (or more) horizon components appear in certain sections. We can resolve the ambiguity when a preferred frame choice is available, as is the case in extreme-mass-ratio mergers.

Finally, note that the horizon within the star can be made to appear arbitrarily early by considering a star of compactness sufficiently close to the black hole limit $\beta= 1/2$. We will find that this can actually happen in the interior of stars that are less compact than a black hole. However, we expect that for physically reasonable interiors the duration of the phenomenon is much more limited, on the order of a few times $M$.  We will return to this point later.

\section{NS-BH mergers in the extreme-mass-ratio limit} \label{sec:nsbhemr}

One might think that the event horizon in a highly dynamical merger can only be obtained with the aid of supercomputer simulations. However, as explained in \cite{Emparan:2016ylg}, in the EMR limit the task becomes enormously simpler. This limit is often taken as one where the size of the large black hole is fixed, while the small object (star or black hole) is regarded as point-like. However, in order to resolve the features of the event horizon in the fusion, we will keep the small object size finite  while the large one grows infinitely large in comparison.

The equivalence principle asserts that we can always place ourselves in the free fall frame of the small object where it is at rest. Since the curvature created by the large black hole is inversely proportional to its size, in the EMR limit this curvature can be neglected in the region around the small object. But the horizon of the large black hole is still present: in this limit it becomes an infinite, Rindler-type, acceleration horizon that extends to infinity as a planar null surface. Therefore, the event horizon of the merger can be found by tracing in the geometry of the small object a family of null geodesics that, far from the object, approach a planar horizon.

This idea was used in \cite{Emparan:2016ylg} to study the merger event horizon when the small object is a non-rotating, Schwarzschild black hole, and in \cite{Emparan:2017vyp} when it is a Kerr black hole. The generalization to other objects is straightforward. Let us consider a spherically symmetric star, with a generic geometry of the type
\beq\label{stargeom}
ds^2=-f(r)dt^2+\frac{dr^2}{g(r)}+r^2\lp d\theta^2 +\sin^2\theta\, d\phi^2\rp\,.
\eeq
In the exterior of a star of radius $R$, Birkhoff's theorem dictates that the geometry must be Schwarzschild's, hence
\beq\label{extsch}
f(r)=g(r) =1-\frac{2M}{r}\,,\qquad r\geq R\,.
\eeq
Naturally, $R> 2M$. The interior geometry depends on the matter model, which we will specify later.

In these coordinates the star is at rest, and $t$, which is a Killing coordinate, measures proper time for the star.\footnote{Viewing this as the first step in a matched asymptotic expansion, \eqref{stargeom} corresponds to the near-zone geometry and $t$ is the near-zone proper time, measured at large $r$.} In this spacetime, we have to find a null geodesic congruence that approaches a null plane at future infinity. This will be the horizon of the large black hole, which, from the viewpoint of the infalling star, is coming towards it and will eventually move through it, leaving the star inside the large black hole. Notice the crucial role that the equivalence principle plays in this, making the phenomenon describable in the free-fall frame of the star while the curvature of the large black hole becomes negligible.

Observe that in this construction (with \eqref{stargeom} describing a static star and not a black hole), the geometry does not have any compact trapped surfaces nor any apparent horizons\footnote{In the EMR limit as we take it, the apparent horizon of the large black hole becomes non-compact. For the EMR binary black hole merger this has been studied recently in \cite{Booth:2020qhb}.}. It is not unusual to find that a compact section of the event horizon exists temporarily within the star without any apparent horizon inside it; in fact, this is common in the collapse of a star to form a black hole, as is easily illustrated with a Vaidya collapsing shell model: although the shell interior is flat space, the event horizon begins to grow there in anticipation of the collapse\footnote{See \cite{Ashtekar:2004cn,Booth:2005qc} for reviews.}. What makes our example perhaps more surprising is that there is no real collapse since all the star matter is static in its rest frame.

\subsection{Equations}

We can now proceed to the construction of the event horizon. As in \cite{Emparan:2017vyp}, we work with the Hamiltonian form of the geodesic equations, which uses the coordinates $x^\mu$ and canonical conjugate momenta $p_\mu$. 
After accounting for the isometries of the collision, the non-trivial equations are
\begingroup
\begin{subequations}
\label{geoeqns}
\begin{align}
  \frac{dt}{d\lambda} &= f(r)^{-1}\,, \label{eq:tdot} \\
  \frac{dr}{d\lambda} &= g(r)\,p_{r}\,,  \label{eq:rdot} \\
  \frac{d\theta}{d\lambda} &= \frac{q}{r^2}\,, \label{eq:thetadot} \\
  \frac{dp_r}{d\lambda} &= - \frac{f'(r)}{2f(r)^2}-\frac{g'(r)}2 p_r^2 +\frac{q^2}{r^3}\,. 
\end{align}
\end{subequations}
\endgroup
Here $\lambda$ is an affine parameter. The geodesics lie on planes of constant angle $\phi$. The integration constant for the energy of the geodesic is normalized to one, and the constant angular momentum $q$ labels the geodesics by their impact parameter at infinity. 

Since we know the exterior geometry \eqref{extsch}, we can expand the equations at large distance $r\gg M$ and large values of $\lambda$, and integrate them analytically order by order. We then fix the integration constants so that the solution asymptotes to the null plane we seek. For the first orders this yields\footnote{We may always add a constant to $t$, \eg to have the pinch at $t=0$, as in fig.~\ref{fig:cuts}.}
\begingroup
\begin{subequations}
\label{eq:consts}
\begin{align}
t(\lambda)&= \lambda +2M \log \lambda -\frac{4M^2}{\lambda}+\frac{M(q^2-8M^2)}{2\lambda^2} +\mathcal{O}(1/\lambda^3)\,,\\
r(\lambda)&=\lambda +\frac{q^2}{2\lambda}-\frac{M q^2}{2\lambda^2}+\mathcal{O}(1/\lambda^3)\,,\\
\theta(\lambda)&=-\frac{q}{\lambda}+\mathcal{O}(1/\lambda^3)\,,\\
p_r(\lambda) &= 1+\frac{2M}{\lambda}-\frac{q^2-8M^2}{2\lambda^2}+\mathcal{O}(1/\lambda^3)\,.
\end{align}
\end{subequations}
\endgroup
These provide the asymptotic conditions at large $\lambda$ for the numerical integration of the geodesic equations in the entire spacetime. Then, by varying $q\in [0,\infty)$ we obtain a one-parameter family of geodesics, which, when rotated along the symmetry axes $\theta=0,\pi$, rule a null 3-surface in the spacetime. Points on this surface can be labelled by $(\lambda, q,\phi)$.

In order to visualize the results, we employ Cartesian-like coordinates
\beq
x= r\sin\theta\,,\qquad z=r\cos\theta\,,
\eeq
and draw the event horizon as a two-dimensional surface in the space $(t,x,z)$. Although $x$ and $z$ do not have separate invariant meaning\footnote{This embedding of the event horizon in three dimensional space is not isometric, but it is simple, intuitive, and illustrative enough for our purposes.}, $\sqrt{x^2+z^2}=r$ does as the area-radius. The full surface is obtained by rotating in $\phi$ around the axis $x=0$. At large $\lambda$, the geodesic congruence approaches the null plane $x= q$, $z= \lambda$, as desired.

\subsection{Star models}

For the interior geometry we have chosen the T-VII model found by Tolman in \cite{Tolman:1939jz} and the Schwarzschild interior solution \cite{Schwarzschild:1916ae}.

T-VII is an analytic solution to the Einstein equations for a perfect-fluid star where the mass density varies quadratically,
\beq
\rho=\rho_0 \lp 1-\frac{r^2}{R^2}\rp\,.
\eeq
The solution for $r\leq R$ is
\beq\label{T-VIIf}
f(r)=\lp 1-\frac{5\beta}{3}\rp \cos^2\lp C -\frac12\ln \lp\frac{r^2}{R^2}-\frac56+\sqrt{\frac{g(r)}{3\beta}} \rp \rp \,,
\eeq
\beq\label{T-VIIg}
g(r)=1-\beta \frac{r^2}{R^2}\lp 5-\frac{3r^2}{R^2}\rp\,.
\eeq
The constant
\beq
C=\frac12 \ln\lp \frac16+\sqrt{\frac{1-2\beta}{3\beta}}\rp+\arctan \sqrt{\frac{\beta}{3(1-2\beta)}}
\eeq
is chosen to ensure the continuity with the metric functions \eqref{extsch} at $r=R$. The solution is scale-free, with only one essential dimensionless parameter, namely the compactness
\beq
\beta=\frac{M}{R}
\eeq
(we are setting $G=c=1$ in \eqref{defbeta}), and where the radius appears in the metric only through $r/R$. In terms of these parameters, the central density is $\rho_0=\frac{15}{8\pi}\frac{\beta}{R^2}$.

The central pressure remains finite as long as  \cite{Lattimer:2000nx}
\beq
\beta\lesssim 0.3862\,,
\eeq
and the sound speed is subluminal when
\beq
\beta \lesssim 0.2698\,.
\eeq
In the latter range, according to  \cite{Lattimer:2000nx} this model provides a good approximation to a wide variety of matter equations of state without exhibiting unphysical behavior. 

The Schwarzschild interior solution \cite{Schwarzschild:1916ae} describes a star made of incompressible fluid. It is also a scale-free solution, with
\beq\label{schintf}
f(r)=\frac14 \lp \sqrt{1-2\beta\frac{r^2}{R^2}}-3\sqrt{1-2\beta}\rp^2\,,
\eeq
\beq\label{schintg}
g(r)=1-2\beta\frac{r^2}{R^2}\,.
\eeq
Its form is simpler than T-VII, but its uniform density is not realistic, in particular because it makes the speed of sound everywhere infinite. Nevertheless, the pressure at the center remains finite whenever the Buchdahl bound \eqref{Buchbound} is satisfied.

Henceforth we fix the mass to
\beq
M=1
\eeq
and then let $\beta=1/R$ be the only parameter characterizing the star.

\subsection{Results}

\begin{figure}
\centering
  \includegraphics[width=.47\textwidth]{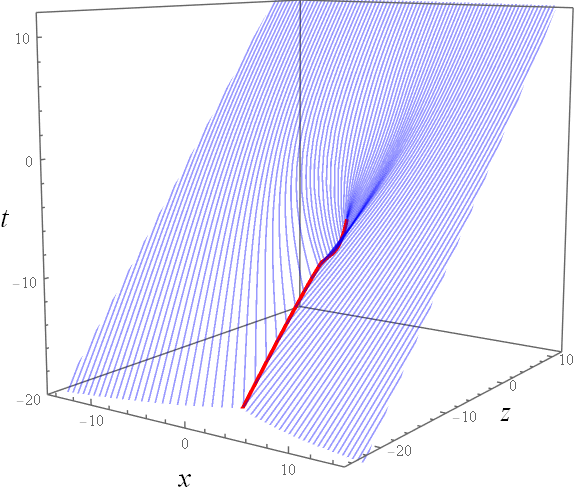}
  \qquad
  \includegraphics[width=.47\textwidth]{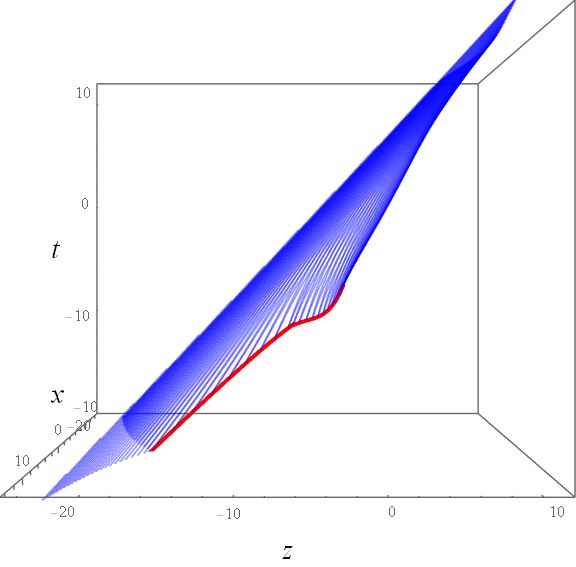}
  \smallskip
\caption{\small Event horizon for the NS-BH merger of a T-VII star with $\beta=0.2$ ($R=5$). The line of caustics is marked in red.}
\label{fig:lobeta}
\end{figure}

\begin{figure}
\centering
  \includegraphics[width=.47\textwidth]{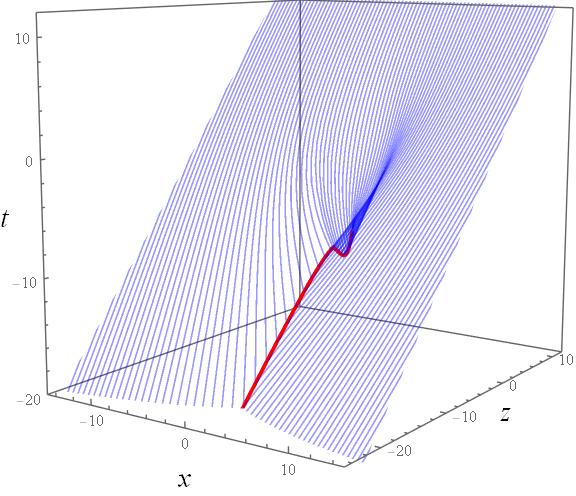}
  \qquad
  \includegraphics[width=.47\textwidth]{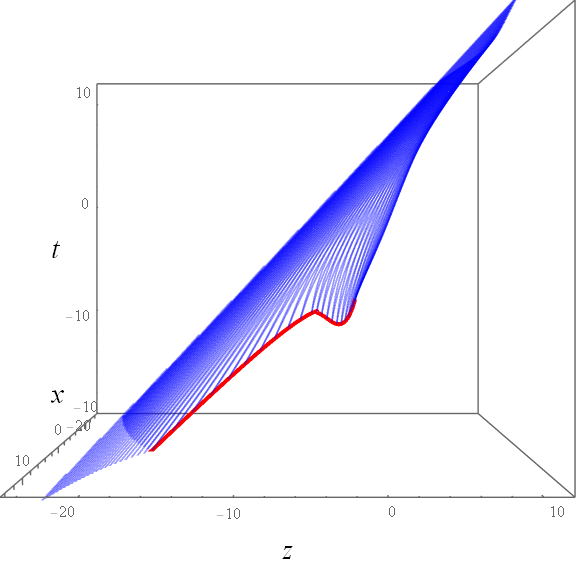}
\smallskip
\caption{\small Event horizon for the NS-BH merger of a T-VII star with $\beta=0.25$ ($R=4$).  The line of caustics is marked in red. This star is compact enough to start its collapse before the merger.}
\label{fig:hibeta}
\end{figure}

We can now plug the interior and exterior metric functions $f(r)$ and $g(r)$ in the geodesic equations \eqref{geoeqns} and numerically integrate them back from large values of $\lambda$ with initial values given by \eqref{eq:consts}. We have done this for stars with a range of values of $\beta=R^{-1}<1/2$. In both star models our results agree with the general picture presented in sec.~\ref{sec:qualia}: for low $\beta$, the caustic line along the axis increases monotonically in the proper time of the star, but for large enough $\beta$, it reaches a maximum at a certain instant, then  decreases until it reaches a minimum, to continue up again for a while until it stops. In figures \ref{fig:lobeta} and \ref{fig:hibeta} we show representative instances of the event horizon in each case for T-VII stars.

The most important parameter to characterize the event horizon is the value of the radius $r_*$ of the saddle point at which the caustic reaches a maximum in $t$. If this saddle exists, then there are two horizon components in constant $t$ slices prior to the merger. The two horizons fuse along the collision axis at the radius $r_*$, in a manner that is captured by a universal, exact local model described in \cite{Emparan:2017vyp}.

When the small object is a Schwarzschild black hole, \cite{Emparan:2016ylg} found the saddle at\footnote{In \cite{Emparan:2016ylg} this was obtained by as the numerical root of an exact transcendental equation. The values for T-VII and Schwarzschild stars that we refer below have been obtained from the numerically constructed event horizons.}
\beq
r_*=r_{*s}\equiv 3.5206\,.
\eeq
By Birkhoff's theorem, if the radius of the star is $R<r_{*s}$, then this saddle point also exists outside the star. In this case precursory collapse occurs, subject only to the approximation of spherical symmetry, independently of the interior geometry. This translates into a model-independent bound on the compactness for precursory collapse in EMR mergers, namely, 
\beq\label{betas2}
\beta>\beta_{s} =\frac1{r_{*s}}=0.2840\,,
\eeq
which yields \eqref{betas}.

This is a sufficient but not necessary condition for precursory collapse. We have found that  for stars with radii $R>r_{*s}$ a saddle can appear at a radius in the interval $r_{*s} < r_* < R$. For the T-VII stars, we find that this occurs whenever the compactness is larger than 
\beq\label{betaT-VIIb}
\beta_{\textrm{T-VII}} =0.22\,.
\eeq

For the most compact but still non-pathological T-VII star with $\beta= 0.2698$, the duration of the precursory collapse from the moment when the horizon first appears within the star, until the fusion with the large black hole,  is
\beq
t_\textrm{coll}=2.05038\,.
\eeq
Above this compactness, causality is violated in the T-VII star since the speed of sound exceeds the speed of light. Nevertheless, if we push to higher $\beta$  the duration of the collapse increases, until we reach $\beta=0.3682$. At this point the central pressure becomes infinite, and the duration of the collapse diverges as the minimum of the caustic drops to $t\to -\infty$.

For the Schwarzschild interior solution \eqref{schintf}, \eqref{schintg}, we find that the precursory collapse occurs for stars more compact than
\beq
\beta_{\textrm{SchInt}}=0.28\,.
\eeq
This is a little lower than the model-independent value \eqref{betas2}, but rather more stringent than $\beta_{\textrm{T-VII}}$. However, the result is less significant since the model is less realistic. When the upper limit $\beta=4/9$ on these solutions is approached, the duration of the collapse diverges, again due to the infinite pressure at the center of the star. It is natural to conjecture that this is a general result: for stars that satisfy the assumptions of Buchdahl's theorem, as $\beta\to\beta_B$ the precursory collapse in the EMR limit begins at $t\to-\infty$. We expect that for finite mass ratios this time will also be finite.

An even simpler model of a `star', even much less realistic but for which the event horizon can be obtained exactly, is a thin shell model with an empty, flat Minkowski interior. In this case, the precursory collapse is only present for \eqref{betas2}.

We have not explored more models, but the fact that T-VII, which stands out among the most physical analytic solutions \cite{Delgaty:1998uy,Lattimer:2000nx,Raghoonundun:2015wga}, allows precursory collapse without needing to force the parameters, leads us to expect that the phenomenon will also be present in other realistic situations.


\section*{Acknowledgments}
We are very grateful to Marina Mart{\'\i}nez for early collaboration in this project. Work supported by ERC Advanced Grant No.~GravBHs-692951, MECD Grant No.~FPA2016-76005-C2-2-P and AGAUR Grant No.~2017-SGR 754.


\end{document}